\def\year{2021}\relax
\documentclass[letterpaper]{article} 
\usepackage{aaai21}  
\usepackage{times}  
\usepackage{helvet} 
\usepackage{courier}  
\usepackage[hyphens]{url}  
\usepackage{graphicx} 
\usepackage{natbib}  
\usepackage{caption} 
\title{The Deepfake Detection Dilemma: A Multistakeholder Exploration of Adversarial Dynamics in Synthetic Media}
\author{
    Claire Leibowicz,\textsuperscript{\rm 1,*}
    Sean McGregor,\textsuperscript{\rm 2,*}
    Aviv Ovadya\textsuperscript{\rm 3,*} \\
}
\affiliations {
    \textsuperscript{\rm 1} The Partnership on AI \\
   \textsuperscript{\rm 2} XPRIZE Foundation, Syntiant Corp  \\
    \textsuperscript{\rm 3} Thoughtful Technology Project \\
    \textsuperscript{\rm *} These authors contributed equally \\
}

\usepackage[framemethod=tikz]{mdframed}

\usepackage{amsmath} 
\usepackage{amssymb} 
\usepackage{enumitem}
\usepackage{multirow}
\usepackage{cite}
\usepackage{endnotes}
\usepackage{array}  
\usepackage{subcaption}

\newcommand{\takeaway}[1]{$\therefore$ \emph{ #1}} 

\newcommand{\sidebar}[1]{
\begin{mdframed}[hidealllines=true,backgroundcolor=black!5]\textbf{Persona Example}: #1\end{mdframed}} 

\newcommand{\newterm}[1]{\textit{#1}}

\newcommand{\term}[1]{#1}

\newcommand{\roughly}{$\mathtt{\sim}$}

\begin{document}

\maketitle

\begin{abstract}
Synthetic media detection technologies label media as either synthetic or non-synthetic and are increasingly used by journalists, web platforms, and the general public to identify misinformation and other forms of problematic content. As both well-resourced organizations and the non-technical general public generate more sophisticated synthetic media, the capacity for purveyors of problematic content to adapt induces a \newterm{detection dilemma}: as detection practices become more  accessible, they become more easily circumvented. This paper describes how a multistakeholder cohort from academia, technology platforms, media entities, and civil society organizations active in synthetic media detection and its socio-technical implications evaluates the detection dilemma. Specifically, we offer an assessment of detection contexts and adversary capacities sourced from the broader, global AI and media integrity community concerned with mitigating the spread of harmful synthetic media. A collection of personas illustrates the intersection between unsophisticated and highly-resourced sponsors of misinformation in the context of their technical capacities. This work concludes that there is no ``best'' approach to navigating the detector dilemma, but derives a set of implications from multistakeholder input to better inform detection process decisions and policies, in practice. 
\end{abstract}

\section{Introduction}


The information ecosystem is comprised of actors that collect, spread, and consume newsworthy or credible information. Increasingly, these actors are disrupted by persons, organizations, and governments aiming to spread misinformation via text, images, and video \cite{verdoliva2020media,Wardle}. Among the negative impacts are increases in violence, non-consensual sexual exploitation, financial loss, and political unrest \cite{adjer, gregory2019}.

Thus, many actors in the global information ecosystem, and in society in general, have an interest in detecting and stopping the spread of problematic content online, including misinformation. Increasingly, these actors look to programmatic tools that check whether source media has been manipulated or synthesized using AI techniques in order to do so. While synthetic media is not inherently harmful or malicious, and can be used for satirical and artistic purposes, such signals are used by tech platforms and others for evaluating the credibility and potential harmful impact of content \cite{Saltz, Bickert, Roth}. Journalists and fact-checkers want detection tools to determine the authenticity of a source video \cite{Leibowicz, Cohen}. Conversely, organizations concerned with documenting systemic abuses around the world want detection technologies to ensure people cannot claim unmanipulated evidence of abuses are dismissed as fake \cite{Gregory}. 

As one of many efforts intended to counteract the challenges that synthetic media presents, \newterm{artifact detection tools and technologies} have been developed by a variety of actors seeking to reduce the harm spawned by malicous synthetic media; for example, Microsoft has partnered with the Reality Defender tool and Google's Jigsaw has created its own tool called Assembler  \cite{Burt, Alba}. These tools can be used to analyze videos, audio, or images to determine the likelihood that they were manipulated or entirely synthesized, without relying on external corroboration or context. We  use the word \newterm{detector} throughout this document to mean `artifact detector tool or technology' although, as we discuss, they are not the only detection options.

While many research results show advances in detector technology, tests on media encountered `in the wild' often show detectors have serious limitations \cite{Leibowicz, Hwang}. Moreover, we face the challenge that a specific detector, or even an entire detection approach combining many detection models, may be compromised if the synthetic media producer actively works to defeat the detectors \cite{neekhara2020adversarial}.

This leads us to the \newterm{detection dilemma}: the more accessible detection technology becomes, the more easily it can be circumvented. As a corollary, we are faced with many challenging questions that impact the practicability of using detection, and the equity in access and outcomes. How should the technical detection community share their tools and techniques, given that we would like them to be effective not just in research but also in the real world? How can we ensure that such techniques, if effective, are accessible to other actors in the global information ecosystem beyond the largest technology companies, including journalists, fact-checkers, and others in civil society?

People working in cryptography, computer security, and fraud detection think in terms of formal games to answer similar questions. In such games, an \newterm{adversary} actively attempts to defeat a \newterm{defender} \cite{petcher2015foundational}. Both parties adapt their tactics in response to the capabilities and actions of their counterparty. Through time the balance of power can shift between parties, but security games rarely reach a definitive end in the real world. It is useful to adapt such frameworks and games to synthetic media detection, a more recent adversarial challenge. However, the synthetic media detection game will eventually end when synthetic media becomes indistinguishable from unmodified video (at least with respect to \newterm{artifacts}). While current indications from the research community suggest we have not reached the point where indistinguishable synthetic video is possible \cite{yang2019exposing}, our analysis should be viewed as capturing a moment in time that requires adaptation as synthetic content gets more sophisticated. New techniques for generating, concealing, and detecting synthetic media are all actively being developed \cite{Jess}. 
Our work therefore aims to explain the roles of various actors active in misinformation generation and detection to provide both insight into the current state of play and into the likely development of capacities as techniques continue to develop and become more broadly known. Increased sophistication and ubiquity of synthetic media will bring with it increased challenges to sharing robust detection tools with the global information integrity community, and therefore mitigating malicious content online.

\def\arraystretch{1.5}
\begin{table*}
\centering
\begin{tabular}{|p{2.5in}|p{2in}|p{2in}|}
\hline
\textbf{Detection context} & \textbf{Forensic Actors} & \textbf{Forensic Actions} \\ \hline

\textbf{Nation-shaking media analysis:} Content which could alter the lives of millions. &
\begin{minipage}[t]{\linewidth}
\begin{itemize}[nosep,after=\strut]
    \item Top media forensics experts
\end{itemize} 
\end{minipage} & 
Extremely detailed examination with sophisticated and potentially customized tools \\ \hline 

\textbf{Suspect media analysis:} Investigations into disinformation, abuse, and criminal conduct. &
\begin{minipage}[t]{\linewidth}
\begin{itemize}[nosep,after=\strut]
    \item Journalists
    \item OSINT investigators
    \item Law enforcement
\end{itemize} 
\end{minipage} &
Close human examination, primarily with standard tools \\ \hline

\textbf{Automation augmented investigations:} Investigations where it is suspected that synthetic media may play a role (e.g. a disinformation campaign). &
\begin{minipage}[t]{\linewidth}
\begin{itemize}[nosep,after=\strut]
    \item Disinformation investigators (e.g. DFRLab, Graphika)
    \item Platform threat intelligence analysts
    \item Disinformation researchers
\end{itemize} 
\end{minipage} &
Tools pull many pieces of media for detection, providing analysis for human review. \\ \hline

\textbf{Flagged content evaluation:} Evaluation of content that has been flagged as suspicious by users for a platform. &
\begin{minipage}[t]{\linewidth}
\begin{itemize}[nosep,after=\strut]
    \item Platforms (e.g., Facebook)
    \item Platform users
\end{itemize} 
\end{minipage} &
Flagging of suspect media by users, which platforms do automated evaluation of (outputs of which may be revealed to users, or impact ranking, moderation, etc.) \\ \hline

\textbf{Full platform evaluation:}
Detection across entire platforms. &
\begin{minipage}[t]{\linewidth}
\begin{itemize}[nosep,after=\strut]
    \item Platforms
\end{itemize} 
\end{minipage} &
Automated evaluation of all platform content, which may just impact metrics, or which may be revealed to users, or impact ranking, moderation, etc.) \\ \hline

\textbf{General public evaluation:} Detection tools provided directly to users, either within a platform or as a separate application or website. &
\begin{minipage}[t]{\linewidth}
\begin{itemize}[nosep,after=\strut]
    \item Everyday people
    \item Non-specialist journalists and other civil society actors
\end{itemize} 
\end{minipage} &
Usage of tools that can take a media item or online account as input and provide an evaluation meant for a non-expert. \\ \hline
\end{tabular}
\caption{\textbf{Detection Contexts}.
A description of detection contexts, some of the key actors that would be involved in each context, and the actions those actors are likely to take in order to achieve their goals. The rows are ordered according to increasing detector model exposure implied by their contexts.}
\label{tab:investigations}
\end{table*}

\subsection{Multistakeholder Input}
Coping with the real world dynamics and impacts of synthetic media requires multidisciplinary input and attention. \citet{Leibowiczdfdc} described multistakeholder governance of the Deepfake Detection Challenge, a 2020 machine learning competition funded by Facebook to build better deepfake detection models. A group of nine experts ranging from computer vision researchers to misinformation trend experts weighed in on the challenge's governance, and in doing so, articulated the need for increased access to detection technologies for journalists, fact-checkers, and those in civil society, as well as the need to attend to the adversarial dynamics of synthetic media detection \cite{Leibowiczdfdc}. 

Building on \citeauthor{Leibowiczdfdc}'s multistakeholder protocol for informing deepfake detection model creation, we consulted a multidisciplinary group of actors from media, civil society, and industry to inform a framework and assessment of the synthetic media detection dilemma; we facilitated three, hour-long workshop sessions with this cohort and offered a three week review period on the initial ideas informing this document. Of the ten individuals consulted, three worked in research and development at global media entities and had expertise in the needs of modern day newsrooms, two were machine learning researchers, one in product management at a large technology company, two in AI and content policy at large technology companies, and the other two individuals were experts in human rights and the global threats of synthetic media and misinformation more generally.

\subsection{Motivations}

Although a detector that is unknown to the purveyors of misinformation is more effective than one that is published publicly, there is a need to facilitate an ecosystem of \newterm{detection technology sharing} that can reduce the negative impacts of synthetic media \cite{Leibowicz}. This will involve difficult choices around information sharing for well-meaning societal actors including technology platforms, academics, journalists, and other detection organizations. In many cases, the best choice for society may be at odds with an organization's particular self-interest, and may require more restrictions---or more openness---than it might naturally be comfortable with. 

The experience of governments attempting to limit the distribution of strong cryptography software indicates that governmental regulation is unlikely to succeed in requiring a detector exposure protocol \cite{EFF}. However, formalizing the current state of play in the detector community can support researchers so that they can best expose artifact detection models in a way that helps address societal needs \cite{leibowiczopen}. Stakeholders who would directly be using these tools to improve societal outcomes, including the media and misinformation research community, have a greater capacity to know whether the models they work with are trustworthy for their intended use cases. This document seeks to help ground the conversation around these goals, providing useful context and frameworks for making sense of malicious synthetic media.

In the following sections, we map the current state of play of synthetic media detection. To our knowledge, this is the first framework informed by multistakeholder input that describes the technical and adversarial dynamics and implications for the technical detection community. We aim to inform recommendations for responsibly deploying detection technologies, and also to explain to media organizations and other non-technical actors using detection technology what can be concluded about deepfake detection in adversarial settings.

We first describe the existing synthetic media actors and explore their technical capabilities. We then develop personas to illustrate these actors and their capabilities. Next, we describe the types of detectors, their exposure levels, and options for defending effectively against synthetic media generators. We use scenarios with the personas to provide further context and implications for different detector exposure levels. Finally, we discuss the key lessons that emerge from this analysis and the implications for the detection dilemma. These involve taking into account not just what detection can do, but other levers and needs, such as app store policies and forensic tool user training, that will enable accessible and responsible detection deployment. 





\section{Synthetic Media Actors and Adversaries}

Grounding this work in operational, real world information contexts requires first detailing the reality of how detectors may be applied. Table \ref{tab:investigations} presents seven detection contexts associated with a variety of stakeholders using detection. These contexts typically have a variety of human processes (e.g., making the determination whether a story should be published or not), but in some cases the detector may make content decisions in a completely automated fashion. These process decisions help determine the exposure of the model to adversaries. A model that is selectively applied to only nation-shaking media can be tightly controlled and applied to a select few pieces of media, while web platforms may apply a detector billions of times. This work focuses on the efficacy of detectors absent human processes to adapt their decisions to evolving adversarial capabilities.

The misinformation actors working to avoid detection in these contexts exhibit a spectrum of capabilities for both synthetic media generation and detection circumvention. We next introduce two classifications describing misinformation actors active in creating and deploying synthetic media technologies, tools, and content.

\subsection{Technical competency spectrum}

The success of actors in evading synthetic media detection depends in part on their technical resources. The spectrum of technical competency can be split into three main categories:

\begin{itemize}
    \item \textbf{Novel resource actors:} Can do original research and implement complex systems from scratch.
    \item \textbf{Common resource actors:} Rely on models and implementations created by novel resource actors. Can build custom tools and pipelines that do not require researcher-level knowledge and understanding. 
    \item \textbf{Consumer resource actors:} Rely on apps and websites created by others. Can only create content and not the tools for creating the content.
\end{itemize}

Most known examples at these sophistication levels are not purveyors of misinformation. The academic creators of a system called \newterm{pix2pix} \cite{pix2pix} would be \textbf{novel resource actors} if they developed the software for disinformation purposes. Pix2pix is a machine learning system for image-to-image translation developed for benign purposes. For example, a demo of pix2pix lets you translate a sketch drawing of a shoe into a photo-like image of a shoe matching that outline, and one can similarly use it to colorize grayscale images or to stylize a photograph \cite{Hesse}.

However, it can also be used for more harmful purposes. ``DeepNude'' is an example of a deepfake generator that pipelines several machine learning advances to make a comprehensive framework for image manipulation. DeepNude is capable of translating photos of people to manipulated images without clothing, and notably does not require the subject's consent \cite{Cole}. While the creators of DeepNude are certainly \textbf{novel resource actors}, their users are \textbf{common resource actors} because the software greatly reduces both the effort and expertise required to generate the deepfake. However, the most realistic results from DeepNude still require understanding the underlying system.

Finally, research advances are likely to enable true \newterm{one-shot image translation}---in other words, the creation of a system which can be given a single image pair, e.g. ``shoe outline $\rightarrow$ shoe'' or ``clothed person $\rightarrow$ unclothed person,'' and then repeat that operation with any future input. A consumer grade tool that lets anyone use this capability could be incredibly valuable to artists and scientists---but would make creating non-consensual unclothed images easy for even \textbf{consumer resource actors.} 
 Fundamental capability advances have many positive applications, but they can also be misused. Ultimately, the degree of technical competence required to create malicious synthetic media is likely to decrease as research results are translated into general purpose consumer products.


\subsection{Anti-detection competence spectrum}

Building on the general notion of technical competence, we are particularly interested here in how competent an actor is at avoiding detection, or at providing tools that help others avoid detection. We can define an anti-detection competence spectrum closely related to the technical competency spectrum:

\begin{itemize}
    \item \textbf{Novel anti-detection actor:} Can develop methods of circumvention that require significant technical resources or researcher knowledge.
    \item \textbf{Determined anti-detection actor:} Can probe detector systems, execute strategies to defeat them, and plumb together pipelines to support this, building on work from novel anti-detection actors.
    \item \textbf{Consumer anti-detection actor:} Can use consumer anti-detection tools.
\end{itemize}

It is not only malicious actors who might be interested in overcoming detection. Curious, status-seeking, and even deeply principled actors who are not proponents of malicious use may intentionally do novel research or development to thwart detection too. For example, an academic or hobbyist may publicize methods on how to overcome detection systems, even if there is no way to address the flaws that they uncovered, under the principle that ``security through obscurity'' is harmful. Of course, they would be ignoring the history of fraud mitigation which often relies on a period of obscurity to succeed. There is a long history of such activity in other domains within cybersecurity and fraud, where the net impact of such developments may be ambiguous. Actors with non-malicious intent may therefore significantly expand the availability of anti-detection technology, thereby unintentionally enabling increased malicious use. In fact, since one of the principal methods for generating synthetic media involves engineering detectors that a neural network model must then defeat, thousands of deep learning engineers are implicitly engaged in defeating detection.


\subsection{Personas} 

\begin{figure*}
    \begin{subfigure}{.45\textwidth}
        \centering
        \includegraphics{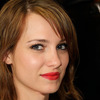}
        \caption{\textbf{Nation-state tech lead Nancy:} a 35-year-old in a secret military intelligence organization who has access to a staff of top-notch machine learning researchers and security experts. If a person is not on her payroll, she can likely coerce through money or extortion anyone living in her country to provide technical assistance. She is in charge of providing tools for the information operators---some of their content is getting detected by platforms and OSINT experts, and she's supposed to make that problem go away. Her users incidentally have access to excellent surveillance imagery on their targets. The detector community does not know someone like Nancy exists, but the general consensus is that there must be a Nancy working for one or more of the major industrialized powers.}
        \label{fig:nationstate}
    \end{subfigure}
    \hfill
    \begin{subfigure}{.45\textwidth}
        \centering
        \includegraphics{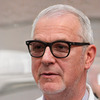}
        \caption{\textbf{Propagandist Paul:} a 65-year-old who uses consumer apps to create emotionally evocative political content for profit---some of which is intentionally manipulated---and which he shares in the 30 online groups he manages, and which occasionally goes viral across many other groups, platforms, and even heavily trafficked partisan news sources. Sometimes when he posts manipulated media that he has created or found, the platform places a warning on the content, or removes it entirely. He sometimes tries to search around to find a way to prevent that from happening, and often compares notes with colleagues of similar viewpoints. He often does just as well remixing and resharing synthesized content created by others.}
        \label{fig:propagandist}
    \end{subfigure}
    \hfill
    \begin{subfigure}{.45\textwidth}
        \centering
        \includegraphics{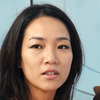}
        \caption{\textbf{Researcher Roberta:} the head of a prominent adversarial machine learning research lab, Roberta is a 38-year-old woman interested in investigating and publishing methods for circumventing synthetic media detectors. She and many of her peers in computer security believe that sharing the details of how to break existing systems is the best way to build more robust systems going forward. As a side effect, her technical work and publicly available source code also makes it easier for unsophisticated adversaries to apply her advanced circumvention techniques, including during a recent spat of political protests in another country.}
        \label{fig:researcher}
    \end{subfigure}
    \hfill
    \begin{subfigure}{.45\textwidth}
        \centering
        \includegraphics{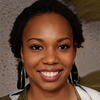}
        \caption{\textbf{Open Ophelia:} a 20-year-old undergrad on summer vacation without an internship, Ophelia is intensely interested in software engineering and artificial intelligence. She doesn't yet understand all the math behind synthetic media generation, but she is eager to learn it so she can produce more realistic digital art. She occasionally reads research papers and closely follows the work of Researcher Roberta. Ophelia plans on integrating Roberta's research code into her favorite open source video editing platform so she can make the most ridiculous (and realistic) TikTok videos ever seen. If/when she does, everyone will be able to make synthetic media with her codebase. }
        \label{fig:open}
    \end{subfigure}
    \hfill
    \begin{subfigure}{.45\textwidth}
        \centering
        \includegraphics{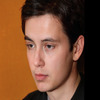}
        \caption{\textbf{Market-manipulator Mike:} works with an irresponsible hedge fund (or perhaps an organized crime 'investment vehicle'), and has similar anti-detection competences to Nation-state Nancy. He may only need to fool automated-trading algorithms for a fraction of a second in order to make hundreds of millions of dollars.}
        \label{fig:markermanipulator}
    \end{subfigure}
    \hfill
    \begin{subfigure}{.45\textwidth}
        \centering
        \includegraphics{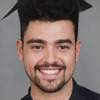}
        \caption{\textbf{Bully Bob:} a teenager who uses consumer apps to synthesize schoolmates saying terrible things in order to embarrass them, get them into fights, or in trouble with the authorities. Bob wants his media to look more realistic but is unwilling to learn how to do anything too complicated. Thanks to the work of Open Ophelia, Bob is starting to get easy access to more advanced techniques. Bob doesn't care if the content is eventually filtered, so long as it is not blocked for a few hours of fun.}
        \label{fig:bully}
    \end{subfigure}
    \caption{\textbf{Personas}. The personas above are fictitious persons rooted in the motivations and capabilities of real persons in the world. They were selected to represent the full range of capabilities and motivations of the misinformation community. Inclusion of prosocial persons (e.g., Researcher Roberta) in this persona list does not constitute classification of those persons as adversaries, but as persons with complex interactions between both the detector community and the purveyors of misinformation. All portraits above are fictitious people and rendered by \emph{ThisPersonDoesNotExist.com}.}
    \label{fig:personas}
\end{figure*}

Personas are fictional persons constructed to represent real persons and are common in user experience engineering and market research. While persona methods are typically used to understand how people interface with a system, we employ them here as a means of fostering collective understanding among diverse stakeholders in the misinformation challenge domain. Figure \ref{fig:personas} introduces six personas that will be referenced in the following sections. These personas are not meant to be exhaustive---there are many additional types of actors---but this is a set found to be useful in stakeholder discussions and covers a wide range of actor capabilities, motivations, and interactions between them. 

\section{Synthetic Media Defenders} 
Having outlined the personas filling rolls in either the generation of misinformation or the required technical infrastructure, this section turns to the ways that one can defend against misuse with detection technology. At present, synthetic media can be detected due to a variety of factors including:
\begin{itemize}

    \item \textbf{Artifacts:} Inconsistencies with the physical world (e.g., strange shifting facial lines, disappearing teeth), or statistical abnormalities (e.g., unnatural audio or video spectrograms). See Bully Bob's persona in Figure 1, part f for an example where Bob's hair blends into what appears to be a graduation cap.

    \item \textbf{Identifiable source material:} The synthetic media is derived from one or more modified source media that can be identified from the post-modification result (e.g., the modified source speech was from a speech delivered six months ago).

    \item \textbf{External context:} Knowledge of synthetic media subjects or environments can highlight inconsistencies (e.g., ``the political candidate was out of the country during the event'' or ``the political candidate has a shoulder impairment and cannot lift his arm above his head'' or ``the sun is not in the right place for the purported time of the year'').

    \item \textbf{Identifiable non-corroborating media:} Other media evidence, such as additional filming angles, can show media as being collectively inconsistent. 

\end{itemize}

These are all factors that can be used to detect synthetic media (i.e., a \newterm{true positive} for a detector), but there are other signals that can establish the veracity of media (i.e., a \newterm{true negative}). In this analysis we focus on the ways that the adversary can fool a detector into reporting a \term{false negative} presuming the content is synthetic. We also specifically concentrate on artifact detection in this work as it is the primary focus of most detection research. However, there is much promising work to be done beyond that, and we encourage additional resources in expanding the utility of complementary approaches. What follows is the state of play at the end of 2020 for synthetic media detection of artifacts with the most common and powerful form of detector: neural networks.

\subsection{Detector Exposure Levels}

Neural networks are biologically-inspired computer programs that network together many chained mathematical operations to perform tasks like classification (e.g., synthetic vs non-synthetic) and synthetic media generation (e.g., transfer a face onto a person). As a tool for both generating and detecting synthetic media, neural networks are of immense interest to the media integrity community. Detector neural networks can be operationally and technically exposed to various degrees to the adversary. Each level of exposure is controlled by the person or organization developing the detector and these choices may include (in order of decreasing exposure):

\begin{itemize}
\item \textbf{Publicly shared trained models:} This gives the public access to all information about the detector, including the ability to run the model themselves. The public then has access to both the full description of the neural network and the configuration (i.e., ``weights'' or ``parameters'' to neural network practitioners).
\item \textbf{Open access queries:} Anyone can make a near-unlimited number of queries to the detector model, to see if the detector considers any number of pieces of content to be synthetic.
\item \textbf{Public untrained models:} The detector model is fully described (e.g., it is published in a public venue) but it is not runnable without appropriate data to configure it (i.e., the network is not trained).
\item \textbf{Controlled access service:} The detector may limit access to querying the detector.
\item \textbf{Private use:} A single organization controls the unpublished model, does not disclose the existence of the detector, and never allows external parties to infer the predictions of the detector.
\end{itemize}

Users of detector models must know the exposure of the model before they can reason about the capabilities of the model to detect synthetic media. In the following sections, we describe each of these exposure levels in more detail and give insight into their implications via persona examples.

\subsubsection{Publicly Shared Trained Model}

Neural network detector models ``learn'' to differentiate synthetic examples from non-synthetic examples by being shown examples of both types of content and iteratively updating their neurons, or weights, as they are known to machine learning practitioners. A similar process is executed when producing a synthetic media generator. However, instead of teaching the neural network to differentiate two types of content, the generator neural network is told to produce content that can fool a second neural network called a \newterm{discriminator} into failing to differentiate synthetic and non-synthetic media. As discriminators improve in their ability to differentiate synthetic and non-synthetic media, the synthetic media generators also improve.

\sidebar{Researcher Roberta publishes a research article showing state-of-the-art performance in synthetic media generation. She achieved the performance by taking all publicly shared trained models and using them to train a new generator. The resulting generator defeats all publicly shared trained models. Open Ophelia doesn't care that the detectors are defeated, but she does want to use the source code to make funny TikTok videos and contribute the generator to her favorite open source video editing project. Now Propagandist Paul, and Bully Bob can both easily use the same open source software to defeat detectors. While the open source software is openly available to them, they often don't go through the trouble since they already work with a different piece of software. However, Nation-state Nancy takes Open Ophelia's source code and integrates it within a growing codebase utilized by the government-funded propaganda teams.

\takeaway{\underline{Novel anti-detection actors} can use new detectors to improve the fidelity of synthetic media.}

\takeaway{If a trained detector neural network is available to the adversary, all \underline{determined adversaries} can hide the artifacts of interest to the detector.}
}

\subsubsection{Open Access Queries}

Detector models can be hosted by organizations like large social networks that then check all user-submitted content for synthetic media. Since the adversary can hide inside a large user population, it is possible to repeatedly submit synthetic media that is repeatedly and imperceptibly altered until it passes through the detector model. This black box attack allows the adversary to defeat a detector that is 99.9 percent accurate by testing, on average, 1,000 variations of the same synthetic media. 

\sidebar{Nation-state Nancy wants a video showing an unfriendly political candidate making bigoted statements to go go viral on social networks. She issues a contract to a security contractor, which generates a deepfake using a publicly available deepfake model. The deepfake generator is already known by the relevant platform, which has a policy against all synthetic media, so it automatically labels the video to users with a warning that the content may be synthetic. Nancy now gives the deepfake to another team in order to remove the warning label. They submit variations of the deepfake with subtle changes to the video (e.g., adding small amounts of random noise) until one or more of the videos is posted without labeling.

\takeaway{If the public can repeatedly test the detector model, determined adversaries will eventually find a way to break it.}
}

\subsubsection{Public Untrained Models}

Researchers studying detector models gain career advancement by publishing their research. Publication norms require at least a full description of what the detector model is doing, but in many cases it is possible to hold back the weights of the neural network from general availability \cite{Campbell}. Without access to the weights, only novel resource actors will be able to train new generator models and detector models based on the knowledge from the research publication. Such actors have the engineering capacity, access to datasets needed for training, and the compute power required to reimplement the detector. 

\sidebar{Researcher Roberta develops a new detector model and publishes it in the proceedings of the International Conference on Synthetic Media Detection (ICSMD). Facebook and Nation-state Nancy immediately begin training secret versions of the model for private use. The datasets amassed by Facebook and Nancy far exceed the datasets available to Roberta, so the resulting detectors are more complete than the one in the academic press. Differences in training mean the resulting networks between Facebook and Nancy have slightly different failings and the detector model at Facebook may detect some synthetic media that passes Nancy's detector model. Still, when Nancy submits synthetic media to Facebook, the chances of synthetic content being detected and labeled as such are greatly diminished.

\takeaway{Publishing a model without training it may significantly increase the effort required to circumvent the model relative to publishing trained models.}

\takeaway{The effectiveness of a detector model depends on the training of the model. The same model may be less effective if trained by actors with less resources.}
}

\subsubsection{Controlled Access Service}

Preventing access to a networked computer program can be difficult, but every detector user can be formally qualified and granted access under controlled circumstances. The detector's exposure may increase if any member of the in group leaks their code or account credentials. A middle ground permitting detector coalitions is to provide outside parties with a fixed number of queries on a detector that is controlled and hosted by the organization producing the detector. Providing the detector implementation is not exfiltrated, access to the detector can be withdrawn from outside parties in the future. When sharing the trained model externally, it is impossible to know whether the outside parties have been compromised.

\sidebar{Only Nation-state Nancy poses a threat to detectors shared via controlled access. She already knows the Google Detector Suite password of the \textit{New York Times} reporter due to a previous operation that installed a keylogger on several reporter laptops. After Nancy uses the access to check whether the misinformation campaign she is about to launch will be detected, Google notices the sudden spike in traffic and sends an email to the \textit{New York Times} user and locks the account. Propagandist Paul remains confused why \textit{The New York Times} is so good at finding and debunking his synthetic media because he doesn't know about the Google Detector Suite.

\takeaway{Trusted parties can share detector results with each other without greatly compromising their effectiveness.}
}

\subsubsection{Private Use}

A detector can potentially be held for private use, meaning no one outside the organization or detector coalition has access or perhaps even knowledge of the detector model's existence. This is the most restrictive use case for the detector model and therefore is the most capable of avoiding active circumvention by the adversary. However, even with such limited exposure there are limitations to what detector models can accomplish. In the case of the DeepFake Detection Challenge (DFDC), teams built their solutions based on a dataset filmed and modified by Facebook, but were ranked for a \$1,000,000 prize purse based on data from the platform's real users \cite{Dolhansky}. The real user data was not created to circumvent detection, but results are not promising as shown in Table \ref{tab:challengedata}.

\begin{table}[]
    \centering
    \begin{tabular}{|c|>{\centering\arraybackslash}p{0.55in}|>{\centering\arraybackslash}p{0.3in}|>{\centering\arraybackslash}p{0.55in}|>{\centering\arraybackslash}p{0.3in}|}
        \hline
         & \multicolumn{2}{c|}{Accuracy} & \multicolumn{2}{c|}{Ranking}  \\ \hline
         & Facebook-Generated Data & User Data & Facebook-Generated Data & User Data \\ \hline
         Competitor 1 & \roughly{}83\% & \roughly{}57\% & 1 & 905 \\ \hline
         Competitor 2 & \roughly{}82\% & \roughly{}65\% & 4 & 1 \\ \hline
    \end{tabular}
    \caption{ \textbf{Two competitor results from the Facebook Deepfake Detection Challenge.} All models degraded significantly from their test set performance on Facebook generated data to test set data defined on user generated data. }
    \label{tab:challengedata}
\end{table}

The top performing competitor on user data has an accuracy of only about 65\% and a performance of 50\% is achievable by flipping a coin. This performance is particularly troublesome in context, where there are millions of non-synthetic videos for every instance of a synthetic video. Thus when a detector is applied to a large population of videos, the likelihood that a video which is labeled as synthetic is actually synthetic is very small. Conversely, a synthetic video generated by a model which was not used to generate training data will likely be missed by the detector.

\sidebar{Twitter develops an information health check dashboard that reports on the rate of synthetic media in user-submitted content. They do not take action on individual accounts (fast system actions constitute ``Open Access Query'' availability), but they do use the information to make community management decisions. The dashboard runs 15 different state-of--of-the--the-art synthetic media detectors and presents the results in a tabular form. After private use of the tool begins, Researcher Roberta publicly publishes a generator that introduces specific people into crowds without swapping their faces. The health check tool does not detect these cases initially, but Twitter eventually adds a 16\textsuperscript{th} detector trained to recognize this specific synthetic media case. Statistically, they know the prevalence of \newterm{crowd injection} generators, but the detector still only classifies two out of three \term{crowd injections} as being synthetic, so it is not used to programmatically label content.

\takeaway{Even when the adversary does not know a detector model exists, the detector will perform poorly on synthetic media types it was not trained to recognize.}

\takeaway{Detector tools cannot currently confirm the integrity of non-synthetic media, but can be one part of a verification process.}
}

\subsection{Defender Signaling Options}

Once a detection system is used to evaluate content, there are a number of actions that can be taken, which give different levels of \newterm{signal} to an adversary about what types of content are detectable:

\begin{itemize}
    \item \textbf{Strongest signal:} Immediate platform actions on the content (remove, label, downrank, etc.).
    \item \textbf{Strong signal:} Direct actions that identify specific detected content. For example:
    \begin{itemize}
        \item Delayed platform actions on the content (may be intentional to weaken the signal, or because of true delays, due to the time needed for human review or even queued automated review).
        \item Calling out of detected content on either social media or established media, by other actors (e.g., civil society, corporate actors, government, individuals).
        \item Offline or off-platform direct action on a creator or distributor, using detected content as evidence.
    \end{itemize}
    \item \textbf{Weak signal:} Delayed response to a user or group, without identifying what piece of content triggered the response. 
    \item \textbf{No signal:} Analytics outputs (tracking extent of such content in aggregate).
\end{itemize}

In addition, if detail is provided about how the content was identified, for example what tools were used, what manipulation was detected, or what signatures were identified, that provides additional information to the adversary.  Of course, not providing that additional context might decrease trust in the claim a piece of media was manipulated or synthesized \cite{encounters}.

\section{Discussion}

There are many obstacles to real world, accessible detection in an adversarial landscape, and detection is only one piece in a much larger suite of problematic content mitigations; yet, detection can act as a useful tool in our malicious content mitigation toolbox. The following summarizes key insights on reckoning with the detection dilemma, building on our analysis above.

\subsection{Aim for ``goldilocks exposure.''}

Keeping detector model sharing closed can clearly be helpful for resisting adversaries---but only up to a point. If detectors are shared too little, then there will be less interest in researching detection and improving the state-of-the-art, most defenders will not get the latest advances, and few if any will keep up with the rate of improvement of synthetic media generation technology.

A norm for academics to avoid publishing state-of-the-art detection models publicly might not prevent the most powerful anti-detection actors with intelligence capabilities that enable them to exfiltrate or recreate such models, but it would prevent novel resource actors like Open Ophelia from making more realistic synthetic media art tools---and thus incidentally prevent Propagandist Paul and Bully Bob from misusing those tools and defeating detectors. If one believes there are many such Pauls and Bobs, then it may be a tradeoff that one chooses to make, even though it may slow innovation and openness. Privately shared datasets are not uncommon in research, and similar (though potentially more rigorously monitored) structures could be used for sharing such models. Similar concerns and approaches apply to models used to generate synthetic media.

\subsection{Every time a detector is used, it becomes less likely to be effective in the future---particularly against novel resource actors.}

In a world where novel resource actors are true threats, the best detectors should be used sparingly, or in ways that provide the least signal to the adversary. One way to do that is with a hybrid model, with detectors of varying levels of quality being provided at different levels of trust and usage. Nation-shaking media, analysis of broad groups, and analytics would utilize the strongest detectors. 

In practice, this might look like sharing the most powerful trained models among a small number of trusted actors developing and deploying detectors. Weaker versions of these detectors could also be provided as a service with rate-limited and monitored access controls to civil society hubs (e.g., via the misinformation fighting non-profit First Draft News), which would then vet users in their networks to give access and the necessary funding required to support their use. The civil society hubs would ideally be incentivized to share credentials as broadly as possible to mitigate gatekeeping---but with strong incentives including threat of loss of access to vet those who are given access thoroughly. If a piece of content satisfied a certain level of potential harm, it would be elevated up the trust chain to better and better detectors. It is also important that alongside access to the technology or tool, such users are granted capacity and training to use them effectively.

Beyond relying on hubs as \newterm{vetting organizations}, there is also the potential to support or create \newterm{media forensics service providers} with powerful tools and resources to support news organizations and other crucial actors. These might be new organizations, or additional programs for existing hubs; they would likely need to be externally funded in order to support under-resourced news and civil society organizations. They would need to not only provide access to tools, but provide training to ensure they can be used effectively and responsibly. 

\subsection{Play offense.}

The less one needs to defend against an adversary, the better. Ideally, all synthesis tools might provide secret signatures, watermarks, consent mechanisms, and so on, including potential Content Authenticity Initiative and Project Origin standards \cite{ovadya2019ethicaldeepfake,origin,CAI}.
Many of these approaches can even be implemented without any decrease in user privacy, though it is vital to think carefully about potential tradeoffs and how such techniques might put individual journalists and rights defenders at risk \cite{GregoryTrust}.

Moreover, it appears to be possible to watermark training data such that the media outputs of the resulting model are also watermarked as synthetic, though there are also definite limitations to this approach requiring more thorough analysis \cite{yu2020artificial}. Since most actors will need to rely on consumer grade tools, those making it easier to create and access such tools have the largest levers. 

It is far easier to simply identify signatures that have been intentionally added to synthetic media by a consumer tool than it is to identify something that may or may not have been created by a tool. Such signatures can also enable higher confidence and enable truly automated responses to synthetic media created through tools that provide signatures. App stores can encourage broad adoption of such signatures by requiring synthetic media tools to utilize them \cite{OvadyaMakingSense}.

These techniques are not capable of defeating Nation-State Nancy, but they will defeat Bully Bob and Propagandist Paul. Where ``perfect is the enemy of the good,'' it is necessary to mitigate rather than solve.


\section{Future Work}
The framework and explanations described above serve as a necessary prerequisite to establishing formal policy, processes, and institutions for detector exposure and collaboration. Having established the actors and capacities of the misinformation ecosystem, civil society organizations and journalists can begin exploring technical collaboration and sharing of tools developed by the largest technology companies. It may be valuable to apply a similar multistakeholder process to the issues of establishing the veracity of media---proving true as opposed to proving false---and the potential adversarial dynamics around that. Even within detection there are many more areas of difficulty which may need to be addressed. For example, adversarial tactics that make it easier to claim doubt or harder to claim certainty instead of attempting to fully defeat detection. There is also the challenge of even defining what counts as synthetic---as almost all media content has some form of synthesis or manipulation---and applying that level of nuance to this analysis. There are some efforts from the fact-checking community, like the MediaReview schema, that are beginning to do this, but there is much work to be done \cite{Benton}. 

However, even if we address the technical side of detection which is the focus of this paper, it is crucial to remember that investment in detection tools will do little if those tools are not sufficiently available, understood, or trusted. Access restrictions to detection tools may help mitigate the detector dilemma in wealthier countries---but insufficient investment in vetting organizations and forensics providers could leave stakeholders in poorer countries with little access. While some might argue that this would be a better outcome than having useless detectors worldwide---a plausible outcome with no access restrictions---it behooves us to aim for equity in detection capacity. Future research should explore how to meaningfully equip stakeholders in less resourced environments to leverage detection technologies.

Moreover, while detection tools may enable content to be flagged on one platform or another, there are plenty of non-platform or closed network methods that nefarious actors have for broadly sharing media. Therefore, in addition to investing in tools, we must resource institutions that enable capacity and legitimacy for detection tools.
This legitimacy can be built on a combination of clear messaging from tool makers and solid governance of the organizations determining who gets access to such tools and technologies. 
In sum, navigating the detection dilemma requires follow up work that evaluates answers to three key questions:
\begin{itemize}
    \item \emph{Who} gets access?
    \item \emph{What} sort of access do these groups or individuals have?
    \item \emph{How} are these groups and individuals chosen?
\end{itemize}
Future work should look to prior instances where these questions have been answered in ways that succeed and fail, and interrogate the reasons why. We need an answer to \emph{why} people and organizations should follow any best practices that emerge.

In our world of democratized innovation, no single actor can truly control detection access and there is no perfect answer for ensuring robustness and access to detection tools. There is simply the work of weighing tradeoffs and then building institutional structures and norms that can continue to navigate those uncertain waters---well enough---through collaboration across disciplines and stakeholders.


\section{Acknowledgments}



Thanks to Sam Gregory for extensive comments on an earlier document draft. Thanks to Emily Saltz, Jonathan Stray, the PAI AI and Media Integrity Steering Committee, and the entire Partnership on AI staff for their input into this project and resultant paper. Thanks to Paul Baranay for \LaTeX{}ing and editing.


\bibliography{base}

\end{document}